\documentclass[journal,12pt,onecolumn,dratclsnofoot,]{IEEEtran} 
\usepackage[mathscr]{euscript}
\usepackage{color}
\usepackage{optidef}
\usepackage[algo2e,ruled,vlined]{algorithm2e}
\usepackage{float}
\usepackage{graphicx}
\usepackage{tikz,pgfplots}
\usepackage{upgreek}
\usepackage{multirow}
\usepackage{yfonts}
\usepackage{mathtools}
\usepackage[normalem]{ulem}
\usepackage{latexsym}
\usepackage{url}
\usepackage{amsmath,amssymb,amsbsy,amsfonts}
\usepackage{mathrsfs}
\usepackage{bbm}
\usepackage[export]{adjustbox}

\newcommand\restr[2]{{
  \left.\kern-\nulldelimiterspace 
  {#1}\vphantom{\big|} \right|_{#2}}}

\newcommand{\R}{\mathbb{R}}

\usepackage{xspace}

\usepackage{cite}
\usepackage{amsmath,amssymb,amsfonts}
\usepackage{graphicx}
\usepackage{textcomp}
\def\BibTeX{{\rm B\kern-.05em{\sc i\kern-.025em b}\kern-.08em
    T\kern-.1667em\lower.7ex\hbox{E}\kern-.125emX}}

\usepackage[math]{cellspace}
\usepackage{graphicx}
\usepackage{algorithm}
\usepackage{algpseudocode}

\usepackage{float}
\usepackage{enumerate}
\usepackage{url}
\usepackage{multirow}

\usepackage{wrapfig,lipsum,booktabs}

\pgfplotsset{compat = 1.17}
\begin{document} 

\title{Technical Note: An Efficient Implementation of the Spherical Radon Transform with Cylindrical Apertures}

\author{Luke Lozenski, Refik Mert Cam, Mark A. Anastasio, and Umberto Villa
\thanks{Luke Lozenski is with Department of Electrical and Systems Engineering,
 Washington University in St$.\,$Louis, St$.\,$Louis, MO 63130, USA}
\thanks{Mark Anastasio is with Department of Bioengineering, University of Illinois Urbana-Champaign, Urbana, IL 61801, USA}
\thanks{ Umberto Villa is with the Oden Institute for Computational Engineering and Sciences, University of Texas at Austin, Austin, TX 78712.}
\thanks{Further author information: (Send correspondence to Umberto Villa.)\\ E-mail: uvilla@oden.utexas.edu, Telephone: +1 512 232-3453}}



\maketitle


\begin{abstract}

The spherical Radon transform (SRT) is an integral transform that maps a function to its integrals over concentric spherical shells centered at specified sensor locations. It has several imaging applications, including synthetic aperture radar and photoacoustic computed tomography. However, computation of the SRT can be expensive. Efficient implementation of SRT on general purpose graphic processing units (GPGPUs) often utilizes non-matched implementation of the adjoint operator, leading to inconsistent gradients in optimization-based image reconstruction methods. This work details an efficient implementation of the SRT and its adjoint for the case of a cylindrical measurement aperture. Exploiting symmetry of the cylindrical geometry, the SRT can then be expressed as the composition of two circular Radon transforms (CRT).  Utilizing this formulation then allows for an efficient implementation of the SRT as a discrete-to-discrete operator utilizing sparse matrix representation. 
\end{abstract}

\section{Introduction}

This technical note details a computationally efficient implementation of the spherical Radon transform (SRT) and its adjoint in the case of a cylindrical measurement geometry. The SRT is an integral transform that describes integrating over a spherical shell in three spatial dimensions (3D) given a specified (sensor) position and radius. The SRT is of particular interest because it can be used to describe acoustic wave propagation with applications to synthetic aperture radar \cite{redding2003inverting} and photoacoustic computed tomography (PACT) \cite{Poudel_2019, burgholzer2007exact, burgholzer2007temporal, zhou2016tutorial, fawcett1985inversion}. 

However, na\"ive implementations of the SRT can be computationally expensive due to the number of points required for computing each of its associated surface integrals. This computational burden greatly hinders the speed of iterative image reconstruction methods in PACT, due to the need for repeated calls of the SRT operator. While computationally efficient GPGPU implementations of the SRT operator exist, these often rely on the use of an approximated (unmatched) adjoint operator\cite{Poudel_2019}. However, this approximation leads to inconsistent gradients when performing optimization-based image reconstruction and therefore requires special treatments within the numerical optimization method \cite{lou2019analysis}. 

This work leverages a theoretical results by  Haltmeier  and Moon \cite{HaltmeierSRT17} that, for certain measurement aperture geometries, allows to formulate the SRT as the composition of two circular radon transforms (CRTs). Specifically, this work describes a computationally efficient decomposition of the SRT and its matched adjoint when a cylindrical aperture geometry is assumed.

\section{Mathematical formulation}

This sections provides a short derivation of the decomposition of the SRT as discussed in Haltmeier and Moon \cite{HaltmeierSRT17}. Let $f \in L^1(\Omega)$ be a function supported over a compact domain $\Omega \subset \R^3$.  Observe that the spherical Radon transform of $f$, with center $\mathbf{r} = (r_1,r_2,r_3) \in \R^3$  and radius $\ell$, can be expressed as the surface integral
$$[\operatorname{SRT} f](\mathbf{r}, \ell)  =  \int_{|\mathbf{r}-\mathbf{r}'| = \ell}  f(\mathbf{r}') dS,$$
where $|\mathbf{r}-\mathbf{r}'| = \ell$ denotes the spherical shell of radius $\ell$ centered at $\mathbf{r}$ and $dS$ denotes the infinitesimal surface area.

\noindent Performing a change of coordinates to spherical coordinates yields

\begin{eqnarray*}
\int_{|\mathbf{r}-\mathbf{r}'| = \ell}  f(\mathbf{r}') dS & = &\int_0^\pi \int_0^{2\pi} f \left( \mathbf{r} + \ell \begin{pmatrix}  \sin \theta \cos \phi \\ \sin \theta \sin \phi \\ \cos \theta \end{pmatrix} \right) \ell^2 \sin \theta d\phi d \theta\\
{} & = & \int_0^\pi \int_0^{2\pi} f (r_1 + \ell \sin \theta \cos \phi, r_2 + \ell \sin \theta \sin \phi, r_3 + \ell \cos \theta)  \ell^2 \sin \theta d\phi d \theta.
\end{eqnarray*} 




\noindent Bringing a copy of $\ell$ outside the integral with respect to $\phi$ gives

\begin{eqnarray*}
[\operatorname{SRT} f](\mathbf{r}, \ell) &=& \int_0^\pi \left(\int_0^{2\pi} f (r_1 + \ell \sin \theta \cos \phi, r_2 + \ell \sin \theta \sin \phi, r_3 + \ell \cos \theta)  \ell \sin \theta d\phi \right) \ell d \theta\\
{} &= & \int_0^\pi  [\operatorname{CRT}^{1,2} f(\cdot,\cdot,r_3 + \ell \cos \theta)](r_1,r_2,\ell \sin \theta) \ell d\theta,
\end{eqnarray*} 

\noindent where $[\operatorname{CRT}^{1,2} f(\cdot,\cdot,r_3')](r_1,r_2,\ell')$ denotes the CRT of $f$ taking in the $(r_1,r_2)$-plane for a fixed height $r_3'$ and radius $\ell' = \ell \sin\theta$.

\noindent Finally, defining the function $f^{1,2}:\R^4 \rightarrow \R$  such that $f^{1,2}(r_1,r_2,r_3',\ell') \mapsto [\operatorname{CRT}^{1,2} f(\cdot,\cdot,r_3')](r_1,r_2,\ell')$ yields

\begin{eqnarray*}
[\operatorname{SRT} f](\mathbf{r}, \ell) &=& \int_0^\pi  f^{1,2}(r_1,r_2,r_3 + \ell \cos \theta, \ell \sin \theta) \ell d\theta\\
{} & = & [\operatorname{CRT}^{3,4}f^{1,2}(r_1,r_2,\cdot,\cdot)](r_3,0,\ell),
\end{eqnarray*}
where, for any given $(r_1,r_2)$,  $[\operatorname{CRT}^{3,4}f^{1,2}(r_1,r_2,\cdot,\cdot)](r_3,0,\ell)$ denotes the CRT of $f^{1,2}$ taken in the $(r_3,\ell)$-plane centered at $(r_3,0)$ with radius $\ell$.

Note that for a fixed $(r_1,r_2),$ the CRT in the first plane $f^{1,2}$ can be computed for several pairs of $(r_3,\ell)$ and reused for the CRT calculation in the second plane. This means that for a group of columnated sensors, with centers $\{(r_1,r_2,r_3^h)\}_{h=1}^{N_h}$ located on a line parallel to the z-axis, SRT calculations can be computed simultaneously and leverage shared elements between each other, thus reducing computational complexity.  

\section{Algorithms}
\subsection{Evaluation of the (forward) spherical Radon transform operator} Approximating the SRT numerically requires a discrete representation of the object function $f$. 
This work consider the case in which $f$ is discretized using piecewise constant (on each voxel) basis; however, the algorithms presented here naturally generalize to other choices of basis functions (such as piecewise linear). Let then $\textbf{F}\in \R^{M_s\times M_s\times M_z}$ be a voxelized representation of the object function $f$ defined on an $M_s\times M_s \times M_z$ Cartesian grid, where $M_s$ denotes the number of voxels in both the $x$ and $y$ dimensions and $M_z$ denotes the number of voxels in the $z$-dimension. The CRT in both the first and second plane, $\operatorname{CRT}^{1,2}$, and $\operatorname{CRT}^{3,4}$) can then be represented as sparse matrices, $\textbf{A}^{1,2}_{CRT}$ and $\textbf{A}^{3,4}_{CRT}$. These discrete representations of the CRT were implemented using the AirToolsII MatLab toolbox \cite{HansenJorgensen2018} and stored as a compressed sparse column matrix\cite{gilbert1992sparse}.  The discrete SRT for a single column of sensors, belonging to a cylindrical aperture, can then be computed using Algorithm \ref{alg:SRT}.

\begin{algorithm2e}\label{alg:SRT}
\SetAlgoLined

\KwIn{A voxelized spatial image $\textbf{F}\in \R^{M_s\times M_s\times M_z}$.} 
\KwIn{A CRT matrix  $\textbf{A}^{1,2}_{CRT} \in \R^{N_l \times M_s^2}$  with center at $(r_1,r_2)$ and radii $\{\ell_l\}_{l=1}^{N_l}$.}
\KwIn{A CRT matrix  $\textbf{A}^{3,4}_{CRT} \in \R^{N_lN_h \times N_lM_z}$    with centers $\{(r_3^h,0)\}_{h=1}^{N_h}$ and radii $\{\ell_l\}_{l=1}^{N_l}$.}
\KwOut{SRT calculations $\textbf{Y}\in \R^{N_h, N_l}$ with centers $\{(r_1,r_2,r_3^h)\}_{h=1}^{N_h}$  and radii $\{\ell_l\}_{l=1}^{N_l}$.}

\SetKwFunction{FMain}{ColumnSRT}
    \SetKwProg{Fn}{Function}{:}{}
    \Fn{\FMain{$\textbf{F}$,   $\textbf{A}^{1,2}_{CRT}$, $\textbf{A}^{3,4}_{CRT}$}}{

    Reshape $\textbf{F} \in  \R^{M_s\times M_s \times M_z} \rightarrow \textbf{F}\in \R^{M_s^2 \times M_z}.$

    Compute CRT in first plane $$\textbf{Y}= \textbf{A}^{1,2}_{CRT}\textbf{F}.$$

    Reshape $$\textbf{Y} \in  \R^{N_l\times M_z} \rightarrow \textbf{Y}\in \R^{N_lM_z}.$$

    Compute CRT in second plane $$\textbf{Y}\leftarrow \textbf{A}^{3,4}_{CRT}\textbf{Y}.$$

    Reshape $$\textbf{Y} \in  \R^{N_l N_h} \rightarrow \textbf{Y}\in \R^{N_l\times N_h}.$$
    
        \textbf{return} $\textbf{Y}.$ 
        }
 \caption{Discrete Spherical Radon Transform for Cylindrical Apertures}
\end{algorithm2e}

Letting $M = M_s^2 M_z$ be the number of voxels, and assuming the standard sampling condition $N_l = \mathcal{O}(M^{1/3})$, Algorithm \ref{alg:SRT} reduces the computational complexity of the SRT from $\mathcal{O}(M^{5/3})$ to $\mathcal{O}(M^{4/3})$ \cite{HaltmeierSRT17}.

 \subsection{Evaluation of the adjoint operator}

Algorithm \ref{alg:SRTadj} provides a matched implementation of the discrete-to-discrete SRT operator defined by Algorithm \ref{alg:SRT}. Here, $\left(\textbf{A}^{1,2}_{CRT}\right)^\dagger$ and $ \left(\textbf{A}^{3,4}_{CRT}\right)^\dagger$ denotes the adjoint (matrix transpose) of $\textbf{A}^{1,2}_{CRT}$ and $\textbf{A}^{3,4}_{CRT}$, respectively.

\begin{algorithm2e}\label{alg:SRTadj}
\SetAlgoLined

\KwIn{SRT calculations $\textbf{Y}\in \R^{N_h, N_l}$ with centers $\{(r_1,r_2,r_3^h)\}_{h=1}^{N_h}$  and radii $\{\ell_l\}_{l=1}^{N_l}$.}
\KwIn{A CRT matrix  $\textbf{A}^{1,2}_{CRT} \in \R^{N_l \times M_s^2}$  with center at $(r_1,r_2)$ and radii $\{\ell_l\}_{l=1}^{N_l}$.}
\KwIn{A CRT matrix  $\textbf{A}^{3,4}_{CRT} \in \R^{N_lN_h \times N_lM_z}$    with centers $\{(r_3^h,0)\}_{h=1}^{N_h}$ and radii $\{\ell_l\}_{l=1}^{N_l}$.}
\KwOut{Discretized object function $\textbf{F}'\in \R^{M_s\times M_s\times M_z}$.}

\SetKwFunction{FMain}{AdjColumnSRT}
    \SetKwProg{Fn}{Function}{:}{}
    \Fn{\FMain{$\textbf{Y}$,   $\textbf{A}^{1,2}_{CRT}$, $\textbf{A}^{3,4}_{CRT}$}}{

    Reshape $$\textbf{Y}\in \R^{N_l\times N_h} \rightarrow \textbf{Y} \in  \R^{N_l N_h} .$$

    Compute CRT adjoint in second plane $$\textbf{F}'\leftarrow \left(\textbf{A}^{3,4}_{CRT}\right)^\dagger\textbf{Y}.$$

    Reshape $$\textbf{F}'\in \R^{N_lM_z} \rightarrow \textbf{F}' \in  \R^{N_l\times M_z} .$$

    Compute CRT adjoint in first plane $$\textbf{F}'= (\textbf{A}^{1,2}_{CRT})^\dagger\textbf{F}'.$$
    
    Reshape $$\textbf{F}'\in \R^{M_s^2 \times M_z} \rightarrow \textbf{F}' \in  \R^{M_s\times M_s \times M_z}.$$

        \textbf{return} $\textbf{F}'.$ 
        }
 \caption{Adjoint of the Discrete Spherical Radon Transform for Cylindrical Apertures}
\end{algorithm2e}

\section{Code availability}
A python implementation of the algorithm described above is publicly available from \cite{Lozenski2024code} under the GPL-3 open-source license. An applications of this library to dynamic PACT imaging can be found in \cite{Lozenski2022code}.

\bibliographystyle{IEEEtran}
\bibliography{local, references} 

\end{document}